\documentstyle[12pt]{article}
\input epsf

\newcommand{\ba}{\begin{array}}
\newcommand{\ea}{\end{array}}

\newcommand{\beq}{\begin{equation}}
\newcommand{\eeq}{\end{equation}}
\newcommand{\bea}{\begin{eqnarray}}
\newcommand{\eea}{\end{eqnarray}}
\newcommand{\beal}{\setcounter{letter}{1} \begin{eqnarray}}
\newcommand{\eeal}{\addtocounter{equation}{1} \end{eqnarray}}
\newcommand{\none}{\nonumber \\}
\newcommand{\req}[1]{Eq.(\ref{#1})}

\newcommand{\larrow}{\,\,\,\,\hbox to 30pt{\rightarrowfill}
\,\,\,\,}
\newcommand{\slarrow}{\,\,\,\hbox to 20pt{\rightarrowfill}
\,\,\,}

\newcommand{\half}{{1\over2}}

\newcommand{\halfu}{{u\over 2}}
\begin{document}
\begin{titlepage}
\renewcommand{\thefootnote}{\fnsymbol{footnote}}
\renewcommand{\baselinestretch}{1.3}
\medskip
\hfill  UNB Technical Report 98-01\\

\begin{center}
{\large {\bf 
THE GEOMETRODYNAMICS OF SINE-GORDON SOLITONS}}
 \\ \medskip 
\medskip

\renewcommand{\baselinestretch}{1}
{\bf
J. Gegenberg $\dagger$
G. Kunstatter $\sharp$
\\}
\vspace*{0.50cm}
{\sl
$\dagger$ Dept. of Mathematics and Statistics,
University of New Brunswick\\
Fredericton, New Brunswick, Canada  E3B 5A3\\
{[e-mail: lenin@math.unb.ca]}\\ [5pt]
}
{\sl
$\sharp$ Dept. of Physics and Winnipeg Institute of
Theoretical Physics, University of Winnipeg\\
Winnipeg, Manitoba, Canada R3B 2E9\\
{[e-mail: gabor@theory.uwinnipeg.ca]}\\[5pt]
 }

\end{center}

\renewcommand{\baselinestretch}{1}

\begin{center}
{\bf Abstract}
\end{center}
{\small
The relationship between N-soliton solutions to the Euclidean sine-Gordon
equation and Lorentzian black holes in
Jackiw-Teitelboim dilaton gravity is investigated, with emphasis
on the important role played by the dilaton in determining the black hole 
geometry. We show how an N-soliton solution can be used to construct 
``sine-Gordon'' coordinates for a black hole
of mass M, and construct the transformation to
more  standard ``Schwarzchild-like'' coordinates. For
 N=1 and 2, we  find explicit closed form solutions to the dilaton
equations of motion in soliton coordinates, and find the relationship 
between the soliton parameters and the black hole mass. Remarkably,
the black hole mass is non-negative for arbitrary soliton parameters.
In the one-soliton case the coordinates are shown to
cover smoothly a region containing the whole interior of the black hole as
well as a finite neighbourhood outside the horizon. A Hamiltonian analysis
is performed
for slicings that approach the soliton coordinates on the interior, and it is
shown that there is no boundary contribution from the interior.   
Finally we speculate on the sine-Gordon solitonic origin of black hole 
statistical mechanics.}
\vfill
\hfill July 6, 1998 \\ \end{titlepage}

\section{Introduction}

Black holes are currently the subject of much research for two main
reasons. First of all, there is a growing body of empirical evidence that
black holes exist in binary systems as well as at the center of most
galaxies\cite{astro}. Secondly, black holes pose fundamental
problems whose resolution will likely provide important clues about the
interface between quantum mechanics and gravity. In particular, 
 the microscopic
origin of the Bekenstein-Hawking entropy of black holes is not fully understood,
despite much recent progress in a variety
of  contexts\cite{stringy,connection,carlip,induced} . The source
of this problem is the fact that from the outside, black holes
are perhaps the simplest, and least complicated objects in the Universe.
They are pure geometry, and according to no-hair theorems proven in the
sixties, tend to settle into highly symmetric configurations with only
a very few externally observable parameters. It is therefore very difficult
to understand where the dynamical modes needed to account for the huge
Bekenstein-Hawking entropy of black holes might reside. 
\par
Dilaton gravity theories in two spacetime dimensions provide useful
theoretical laboratories for studying these questions. They are
diffeomorphism invariant theories that generically do have black hole
solutions, and yet are simple enough to be exactly solved at both the
classical and quantum levels. One such theory of particular interest is
the so-called Jackiw-Teitelboim theory\cite{jt}, which was originally put forward
because of its connection to the
 Liouville-Polyakov action. Jackiw-Teitelboim gravity
theory is distinguished from other dilaton gravity theories in part because it has a great deal of symmetry. It has a gauge
theory formulation, and its solutions correspond to
 maximally symmetric, constant curvature
metrics in two space-time dimensions.  The black hole solutions to
Jackiw-Teitelboim gravity  are related
by dimensional reduction to the BTZ black hole solutions\cite{btz} of anti-DeSitter
gravity in 2+1 dimensions. As shown in Ref.\cite{lemos}, Jackiw-Teitelboim
black holes exhibit the usual thermodynamic properties, including black hole
entropy, despite the absence of field theoretic dynamical modes in the theory.
\par
The feature of Jackiw-Teitelboim gravity most relevant to 
the present analysis is the 
fact that the black hole solutions are space-times of constant curvature.
The relationship between Euclidean, constant curvature metrics in two
dimensions and Lorentzian sine-Gordon solitons has been appreciated by
mathematicians for a long time\cite{early}.
\footnote{It was also used\cite{dolan}
to derive the relationship between solutions to the Liouville equation
and the sine-Gordon solitons.} In particular,
the solutions of the sine-Gordon 
equation
\beq
-\partial_t^2\phi+\partial_x^2\phi =m^2\sin{\phi},\label{sg}
\eeq
determine Riemannian geometries with constant negative curvature $-2m^2$ whose metric 
is given by the line-element
\beq
ds^2=\sin^2\left({\phi\over2}\right)dt^2+\cos^2\left({\phi\over2}\right)dx^2.
\eeq
The angle $\phi$ describes the embedding of the manifold into a three 
dimensional Euclidean space. \cite{early}
 Moreover, it has recently been proved that
there is a direct connection between genus and soliton number\cite{boya}.
 In a recent letter\cite{soliton1} the
relationship between Euclidean sine-Gordon solitons and black holes in
Jackiw-Teitelboim gravity was derived. This relationship is interesting
in part because sine-Gordon theory has a rich (and well-studied)
dynamical structure, while, as mentioned above Jackiw-Teitelboim gravity
has virtually no dynamical structure. Thus, the question arises as to 
whether one can somehow understand the apparently rich dynamical structure
of black holes in terms of the sine-Gordon solitons.
In this paper, we continue the analysis of \cite{soliton1}, and present new results.  In 
particular,
 we show how an N-soliton solution can be used to construct 
``sine-Gordon'' coordinates for a black hole
of mass M, and show generally
 how to transform to more standard ``Schwarzchild-like'' coordinates. For
 N=1 and 2, we are able to find explicit closed form solutions to the dilaton
equations of motion in soliton coordinates, and find the relationship 
between the soliton parameters and the black hole mass. Remarkably, the black 
hole mass is non-negative for all soliton parameters.
\par

The paper is organized as follows: Section 2 reviews Jackiw-Teitelboim gravity
and describes the corresponding black hole solutions. Section 3 describes
how Euclidean sine-Gordon solitons emerge from Jackiw-Teitelboim gravity, and
constructs the general coordinate transformation that relates ``soliton
coordinates'' to the more usual black hole coordinates. Section 4 adapts
the formalism of Babelon and Bernard\cite{bb} to the
case of {\it Euclidean} N-solitons
and presents the results for the 1- and  soliton sectors. Section 5
explicitly displays the black hole geometries associated with 1- and 2- 
soliton solutions derived in Section 5. Section 6 presents the Hamiltonian 
analysis for slicings that approach the soliton coordinates on the interior, and it is
shown that there is no boundary contribution from the interior. Finally, in
Section 7 we close with conclusions, speculations and prospects for future
work.
\par

\section{Jackiw-Teitelboim Gravity}

Since the solutions of the sine-Gordon equation determine metrics with 
constant negative curvature, we need to consider black holes of this type.  
It is therefore natural to examine black holes in Jackiw-Teitelboim 
gravity \cite{jt}.  This theory, like all local two dimensional 
gravity theories, is not purely metrical;  rather there is, besides the 
metric tensor of spacetime, a real-valued scalar field, called the 
{\it dilaton field}.  

The action functional for Jackiw-Teitelboim gravity is 
\beq
I_{JT}[\tau,g]={1\over2G}\int_{M_2}d^2x\sqrt{\mid g\mid }
\tau\left(R+2m^2\right).
\label{JT}
\eeq
In the above, the spacetime metric is $g_{\mu\nu}$,
$R$ is its scalar Ricci curvature and $g$ is its determinant;  $\tau$ is the dilaton field;  
the constant $m$ is related to the `cosmological constant' $\Lambda$ 
by $\Lambda=m^2$.  Finally, $G$ is the gravitational coupling constant, which in two dimensional
spacetime is dimensionless.  Sufficient conditions that this
functional be stationary under arbitrary variations of the dilaton and
metric fields are, respectively
\bea
R+2m^2=&0;\label{const_cur}\\
\left(\nabla_\mu\nabla_\nu-m^2g_{\mu\nu}\right)\tau=&0.\label{jteqms}
\eea
As shown in \cite{obs}, for every solution $\{g_{\mu\nu},\tau\}$ to
the above field equations there is a Killing vector, $k^\mu$ which
leaves both the metric and the dilaton invariant. It is:
\beq
k^\mu={\epsilon^{\mu\nu}\over m\sqrt{-g}}\partial_\nu\tau,
\label{kv_1}
\eeq
where $\epsilon^{\mu\nu}$ is the permutation symbol.
Moroever, up to diffeomorphisms there exists only a one parameter family
of solutions. This parameter, which we call the
mass observable $M$,  is the analogue of the ADM mass
in general relativity: it is the conserved charge associated with translations
along the Killing direction $k^\mu$, and can be expressed in coordinate
invariant form as\cite{obs}:
\beq
M=-{1\over m^2}\mid\nabla\tau\mid^2+\tau^2
,\label{adm_1}
\eeq

Although all the solutions of Jackiw-Teitelboim gravity are {\it locally}
diffeomorphic to two-dimensional anti-DeSitter spacetime, one may obtain
distinct {\it global} solutions, some of which display many of the
attributes of black holes \cite{obs,robb1}.  For example, consider a solution
where the metric is given by 
\bea
ds^2=-\left(m^2 r^2-M\right)dT^2+\left(m^2 r^2-M\right)^{-1}dr^2,\label{bh}
\eea
and dilaton field by:
\beq
\tau = c_1mr
\label{eq:dil1}
\eeq
This solution corresponds to
standard ``Schwarzschild-like'' coordinatization of the black hole solution to Jackiw-Teitelboim
gravity.  Note that $c_1$ is an arbitrary
constant of integration that  has no direct physical significance: the only
true observable is $M$ as defined above.
The constant $c_1$  can be fixed by imposing suitable boundary 
conditions on the fields. For example, requiring that the dilaton 
go to the vacuum configuration $\tau=mr$ as $r \to \infty$
fixes $c_1=1$. Thus without loss of generality
we henceforth make this choice.
\par
  Clearly there is an event horizon located
at $r=\sqrt{M}/m$.  It is important to note here that though this fact
can be easily read off from the metric, since the latter is in manifestly
static form, it also follows from solving for the variable $r$ in the
equation $\mid k^\mu\mid^2:=g_{\mu\nu}k^\mu k^\nu=0$.
The global structure of the black hole spacetime is in part determined
by the dilaton. Since the spacetime has constant curvature, there are no
curvature singularities for any value of $\tau$. However,  surfaces
for which $\tau=0$ give rise to an infinite effective Newton's constant
and should therefore be excluded from the manifold. With this
assumption, the global structure of the manifold is virtually identical
to the $(r,t)$ section of a Schwarzschild black hole, with exactly the
same Penrose diagram\cite{lemos}. 

One can further  motivate the exclusion of $\tau=0$ surfaces from
the manifold by noting that the  metric \req{bh} is the dimensionally truncated spinless BTZ black hole\cite{btz} in 2+1 anti-De Sitter gravity. The dilaton corresponds to the `missing' 
radial coordinate of the 2+1 solutions. 
As described in \cite{btz}, there  is a causal
singularity  in
the BTZ black hole at $\tau=0$. By excluding these surfaces one removes the
possibility of closed timelike curves. Of course, in the context of JT gravity there
is no causal  singularity. The surface $\tau=0$ is completely regular.

As shown in Section 6, the ADM energy of the
black hole solution \req{bh}, \req{eq:dil1} is 
\beq
E_{BH}=mM/2G,
\eeq
It is straightforward to derive the thermodynamic properties\cite{obs, lemos}
 of
 the black holes
described by \req{bh} and \req{eq:dil1}. The Hawking temperature is:
\beq
T_H= {\sqrt{M}m\over 2\pi },
\eeq
with associated Bekenstein-Hawking entropy:
\beq
S_{BH}= {2\pi \sqrt{M} \over G}.
\eeq
\par
It is important to note that
since the constant curvature metrics are maximally symmetric, there
are three Killing vector fields. This also follows directly from
the dilaton equations of motion in that there exist three
functionally independent solutions $\tau_{(i)}, i=1,2,3$ of
\req{jteqms}, which in turn determine three functionally 
independent vector fields $k^\mu_{(i)}$ via
\beq
k^\mu_{(i)}={\epsilon^{\mu\nu}\over m\sqrt{-g}}\partial_\nu\tau_{(i)},
\label{kv}
\eeq
  These three
vector fields satisfy the Killing equations by virtue of \req{jteqms},
and also leave their respective generating dilaton fields invariant.
(i.e. $k^\mu_{(i)}\nabla_\mu \tau_{(i)}=0$).

 It is straightforward to show that in addition 
to $\tau_{(1)}= mr$
the following are solutions to the dilaton equations (\ref{jteqms})
are:
\bea
\tau_{(2)} &=& \sqrt{m^2r^2 - M}\sinh{\sigma},
\label{eq:dil2}\\
\tau_{(3)} &=& \sqrt{-m^2r^2 + M}\cosh{\sigma},
\label{eq:dil3}
\eea
where $\sigma:= m\sqrt{M}T$ and we have set the overall scale factors
$c_{i}$ to unity.
Note that $\tau_{(i)}$ are functionally independent on non-trivial domains of spacetime.   
The  corresponding
Killing vector fields are:
\bea
{\vec k}_{(1)} &=& (1,0),\\
{\vec k}_{(2)} &=& \left({mr\over\sqrt{m^2r^2-M}}\sinh\sigma,
  -\sqrt{M}\sqrt{m^2r^2-M} \cosh\sigma\right),\\
{\vec k}_{(3)} &=& \left({mr\over\sqrt{-m^2r^2+M}}\cosh\sigma,
  \sqrt{M}\sqrt{-m^2r^2+M} \sinh\sigma\right).
\eea

The conserved charge associated with each solution is:
\beq
M_{(i)}=-{1\over m^2}\mid\nabla\tau_{(i)}\mid^2+\tau_{(i)}^2
,\label{adm}
\eeq
When the corresponding Killing vector is timelike, it can be shown that this corresponds
to the ADM energy of the solution.
In fact, a straightforward calculation shows that the conserved charges \req{adm}
for ${\vec k}_{(i)}$ are all equal:
\beq
M_{(i)}= M.
\eeq

\section{From Sine-Gordon Solitons to Black Holes}
Suppose one  wants to solve the field equation \req{const_cur} 
with metrics of the form:
\beq
ds^2=-\sin^2{\halfu}dt^2+\cos^2{\halfu}dx^2,
\label{sGmetric}
\eeq
It is straightforward to show
that this metric has constant negative curvature $R=-
2m^2$ if and only if $u$ satisfies the {\it Euclidean} sine-Gordon equation
\beq
\Delta u=m^2\sin{u},\label{sG}
\eeq
where $\Delta:=\partial^2_t+\partial_x^2$.
Moreover, for such a metric, the dilaton equations take the following
form:
\bea
\tau''+ {\sin(u/2)\over2\cos(u/2)}u' \tau'
   +{\cos(u/2)\over2\sin(u/2)}\dot{u} \dot{\tau} - {m^2\over2}
\cos^2(u/2)\tau &=&0,
 \label{dxx}\\
\ddot{\tau}-{\sin(u/2)\over2\cos(u/2)}u' \tau'
   -{\cos(u/2)\over2\sin(u/2)}\dot{u} \dot{\tau} + {m^2\over 2}
 \sin^2(u/2)\tau 
 &=& 0,
\label{dtt}\\
\dot{\tau}'- {\cos(u/2)\over2\sin(u/2)}u'\dot{\tau} +
   {\sin(u/2)\over2\cos(u/2)} \dot{u} \tau' &=&0.
\label{dxt}
\eea 
By taking a sum of \req{dxx} and \req{dtt} one finds that the dilaton
$\tau$ must satisfy the linearized sine-Gordon equation:
\beq
\ddot{\tau}+ \tau'' = m^2 \cos(u) \tau.
\label{lsg}
\eeq
Thus as first noted in \cite{soliton1}, the dilaton, which generates
the Killing vectors (i.e. symmetries) of the black hole metric, also
maps solutions of the sine-Gordon equation onto other solutions. That is
if $u$ and $\tau$ obey \req{sG} and
\req{lsg}, then  the
field $u'= u+\epsilon\tau$,  also solves \req{sG} to first order in $\epsilon$.
\par
Another method for deriving the linearized sine-Gordon equation for the
dilaton was presented in \cite{soliton1}, where it was noted that putting
the metric {\it ansatz} \req{sGmetric} directly into the action \req{JT}
yields an action of the form:
\beq
I_{JT}[\tau,u] = {1\over 2G} \int_{M_2}d^2x \,\,\tau (\Delta
 u - m^2 \sin{u}).
\label{reduced}
\eeq
Varying the above with respect to $u$ and $\tau$ yields the linearized
sine-Gordon equation for $\tau $ and the sine-Gordon equaiton for $u$, respectively. It should however be remembered that varying the action after imposing
a metric {\it ansatz} does not necessarily yield exactly the same space of
solutions as obtained when the {\it ansatz} is substituted directly 
into the equations of motion.
There are more solutions to the linearized sine-Gordon equation
than there are to the dilaton equations.  However, it may, under certain 
circumstances be desirable to consider the reduced action \req{reduced} as 
defining the physical theory.  This would be analogous to how the Nordstrom 
theory of gravity is obtained from general relativity in 3$+$1 dimensions 
by requiring that the metric be conformally flat\footnote{We are grateful to 
M. Ryan for pointing this possibility out to us.}.  
We will consider the full set of equations \req{jteqms} as defining our 
theory and not consider this alternative formulation further here.

\par
Given a 
dilaton field $\tau(x,t)$ satisfying the dilaton equations of motion \req{jteqms}, we can 
choose a new `radial coordinate' 
\beq
r(t,x):=\tau/m. \label{r}
\eeq
If this is substituted into \req{sGmetric} and the square in the terms in 
$dt^2$ and $dtdr$ is completed, the metric becomes:
\beq
ds^2=-{\mid\nabla\tau\mid^2\over m^2}dT^2+{m^2\over\mid\nabla\tau\mid^2}dr^2,
\label{bhgen}
\eeq
where, as anticipated by the notation, the differential form
\bea
dT&:=&\tan{\halfu}{\tau_{,x}\over\mid\nabla\tau\mid^2}dt+
 \cot{\halfu}{\tau_{,t}\over\mid\nabla\tau\mid^2}dx\\
&=&{- m\over\mid\nabla\tau\mid^2}*d\tau 
,\label{dT}
\eea
where $*$ is the Hodge dual, is closed in any region where 
the coefficients of $dt,dx$ are smooth.  The latter is a consequence of 
the fact that $\tau$ satisfies \req{jteqms}.  Indeed, rewrite \req{dT} 
as
\beq
T_\mu={\eta_\mu{}^\nu\nabla_\nu\tau\over\mid\nabla\tau\mid^2}
   ={k_\mu\over  m \mid k \mid^2},
\eeq
where $\eta^{\mu\nu}:=\epsilon^{\mu\nu}/\sqrt{\mid g\mid}$ is the completely 
skew-symmetric {\it tensor} in two dimensions.  Now use the fact that 
$\nabla_\mu\eta^{\nu\pi}\equiv 0$ and the dilaton equations of motion \req{jteqms} 
to show that 
\beq
\eta^{\mu\nu}(\partial_\nu T_\mu)=\eta^{\mu\nu}\nabla_\nu T_\mu \equiv 0.
\eeq

Finally using \req{adm}, we can 
write the metric \req{bhgen} in the form of the black hole metric \req{bh}.

It is clear that the sine-Gordon coordinates are singular at $u(t,x)=n\pi, n=0,\pm 1, \pm 2,...$ since the volume element $\sqrt{|g|}$ vanishes at those
space time points. For a generic soliton solution $u$, these coordinate
singularities occur either at the soliton locations ( $n$ odd) or at
spatial infinity, where the soliton solution settles down to its asymptotic
value($n$ even). On the other hand,
 the sine-Gordon coordinates are regular at the
black hole event horizon where $\mid k\mid^2=-\mid\nabla
\tau\mid^2=0$. Since the black hole coordinates are singular at the
horizon,  the transformation from the sine-Gordon coordinates 
$(t,x)$ to black hole coordinates $(T,r)$ breaks down there
(cf. \req{dT}).

\section{Multi-Solitons}

It is well-known that the sine-Gordon equation is integrable, and various 
techniques are available for extracting explicit solutions.  Here we use 
Hirota's method to more-or-less explicitly display the multi-soliton 
solutions of the {\it euclidean} sine-Gordon equation.

Our approach here follows that of Babelon and Bernard \cite{bb}.  In 
light-cone coordinates $z_\pm:=x\pm t$, the Lorentzian signature sine-Gordon 
equation is $4\partial_+\partial_- u=m^2 \sin{u}$, where $\partial_\pm:=
\half(\partial_x\pm\partial_t)$.  We switch to the euclidean signature 
via $t\to it$ and $z_\pm\to\half(\partial_x\mp i\partial_t)$.  The Hirota 
functions $\tau_\pm$ are related to the real function $u$ 
in the sine-Gordon equation 
by
\beq
{\tau_-\over\tau_+}=e^{i u/2}.\label{tau}
\eeq
The Hirota functions 
satisfy the {\it Hirota equations}:
\beq
\tau_\pm(\partial_-\partial_+\tau_\pm)-(\partial_-\tau_\pm)(\partial_+\tau_\pm)
={m^2\over16}(\tau_\pm^2 - \tau_\mp^2).\label{hirota}
\eeq
It is easy to see from \req{tau} that the difference of the 
two Hirota equations 
implies the sine-Gordon equations in light-cone coordinates.  

An N-soliton solution of the sine-Gordon equation is given by
\beq
\tau_\pm^N:=\det(1\pm V^N),\label{tauN}
\eeq
where $V^N$ is the $N\times N$ matrix with elements $V^N_{ij}$ given by
\beq
V^N_{ij}:=2{\sqrt{\mu_i\mu_j}\over\mu_i+\mu_j}\sqrt{X_iX_j},\label{V}
\eeq
where the $X_i$ are 
\beq
X_i:=a_i\exp{\half m(\mu_i z_+ + \mu_i^{-1}z_-)}.\label{X}
\eeq
In the above the $\mu_i$ are complex parameters of modulus unity  and 
the $a_i=\pm i e^{w_i}$, where the $+(-)$ sign signifies a soliton 
(anti-soliton) and the $w_i$ are real.  In fact, the $w_i$ can be `absorbed' 
into the exponent of the $X_i$ by writing $w_i:=-\mu_i\xi_+-\mu^{-1}_i\xi_-$ 
and rewriting $z_\pm\to z_\pm-\xi_\pm$.  The reality conditions on the 
parameters are required so that $u$ is a real-valued solution of 
the euclidean sine-Gordon equation.  For the Lorentzian sine-Gordon equation, 
the $\mu_i$ are real.

It is useful to redefine the parameters as follows.  Write $\mu_i=\cos{\beta_i}
+i\sin{\beta_i}$.  Then define $v_i:=\tan{\beta_i}$, so that $\gamma_i:=
1/\sqrt{1+v^2_i}=\cos{\beta_i}$.  In this case the $X_i$ can be written as
\beq
X_i:=i\epsilon_i e^{\rho_i},\label{newX}
\eeq
where
\beq
\rho_i:=m\gamma_i[x-x_0^i-v_i(t-t_0^i)],\label{rho}
\eeq
with $w_i=m\gamma_i(x_0^i-v_i t_0^i)$ and $\epsilon_i=\pm 1$.

Using this notation, the well known 1-soliton solution is obtained from:
\bea
e^{-iu/2}=\cos{u/2}-i\sin{u/2}&=&{\tau_+\over\tau_-}\\
&=&{1+i\epsilon e^\rho\over 1-i\epsilon e^\rho},
\eea
which yields:
\beq
u=4\tan^{-1}\pm e^{(\rho)}.\label{1sol_1}
\eeq
\par
For the 2-soliton, the Hirota functions are
\beq
\tau_\pm=1\pm(X_1+X_2)+\lambda^2
X_1X_2,
\eeq
where
\beq
\lambda:=\left({\mu_1-\mu_2\over\mu_1+\mu_2}\right).
\eeq
Now the solution $u$ of the sine-Gordon equation is given by
\bea
e^{-iu/2}={1+\lambda^2X_1X_2+(X_1+X_2)\over 1+\lambda^2X_1X_2-(X_1+X_2)}.
\eea

In terms of the more `physical' parameters $v_1,v_2$, we write $\lambda
=i\ell$, with $\ell$ real and given by 
\beq
\ell={\gamma_1\gamma_2(v_1- v_2)\over 1+\gamma_1\gamma_2(1+v_1 v_2)}.
\label{lambda}
\eeq
From this it follows that
\beq
u=4\tan^{-1}\left|\epsilon_1 e^{\rho_1}+\epsilon_2 
e^{\rho_2}\over 1+\ell^2\epsilon_1\epsilon_2
e^{\rho_1+\rho_2}\right|.\label{2-sol1}
\eeq
In the case $\epsilon_1\epsilon_2<0$, $u$ describes an $N=2$ soliton which behaves asymptotically
as two 1-solitons and may be viewed as the scattering of the 1-solitons
form each other. For $\epsilon_1\epsilon_2>0$, on the other hand,
 \req{2-sol1} describes a soliton-anti-soliton scattering solution.

\par
\begin{figure}[v]
\vspace*{-3cm}
\epsfxsize=16 cm
\epsfbox{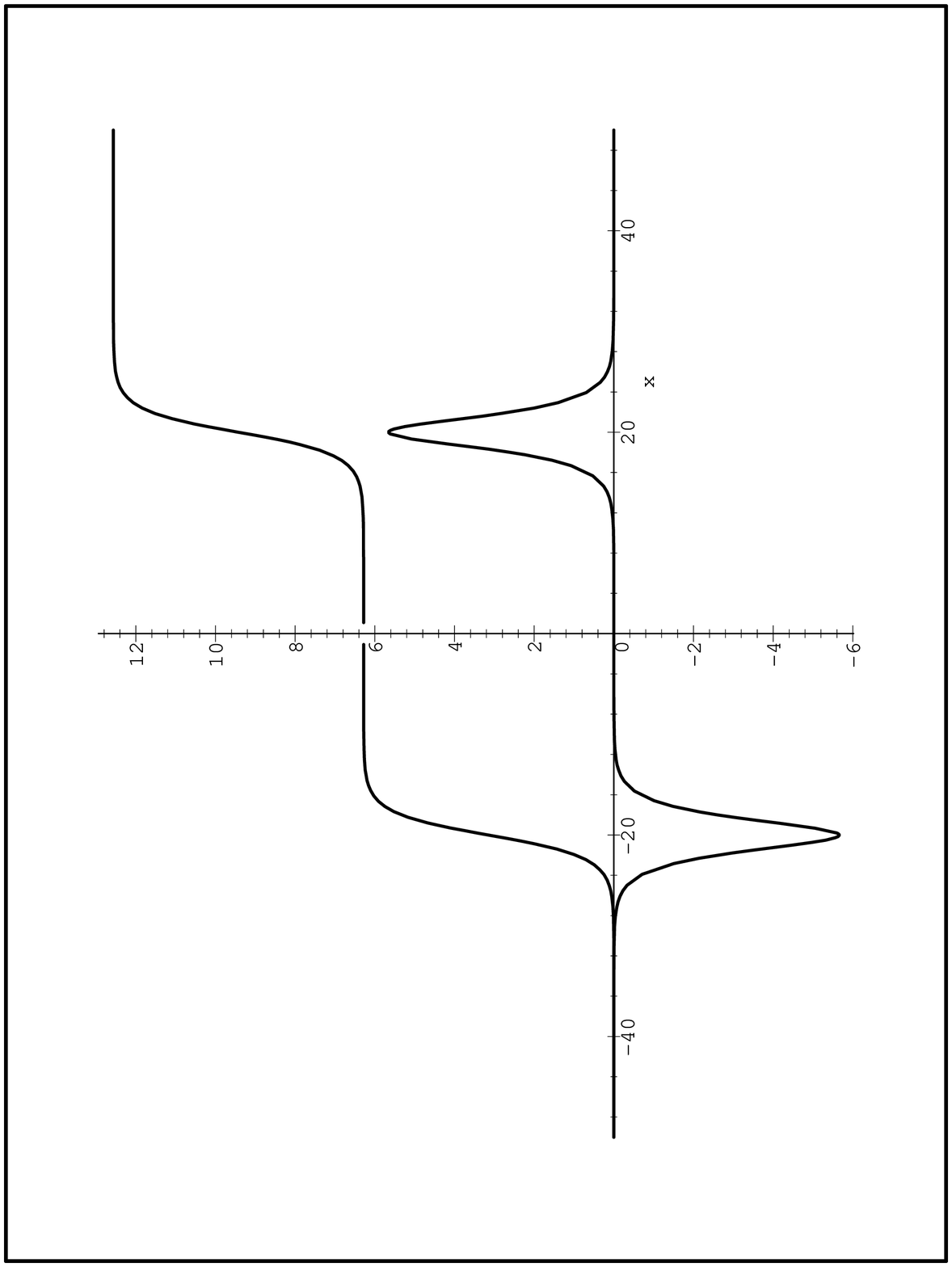}
\caption{Graph of soliton-soliton solution (solid line) and corresponding dilaton at (dashed line)
fixed t (not to scale).}
\label{2soliton}
\end{figure}
\begin{figure}[hbt]
\vspace*{-3cm}
\leavevmode
\epsfxsize=16 cm
\epsfbox{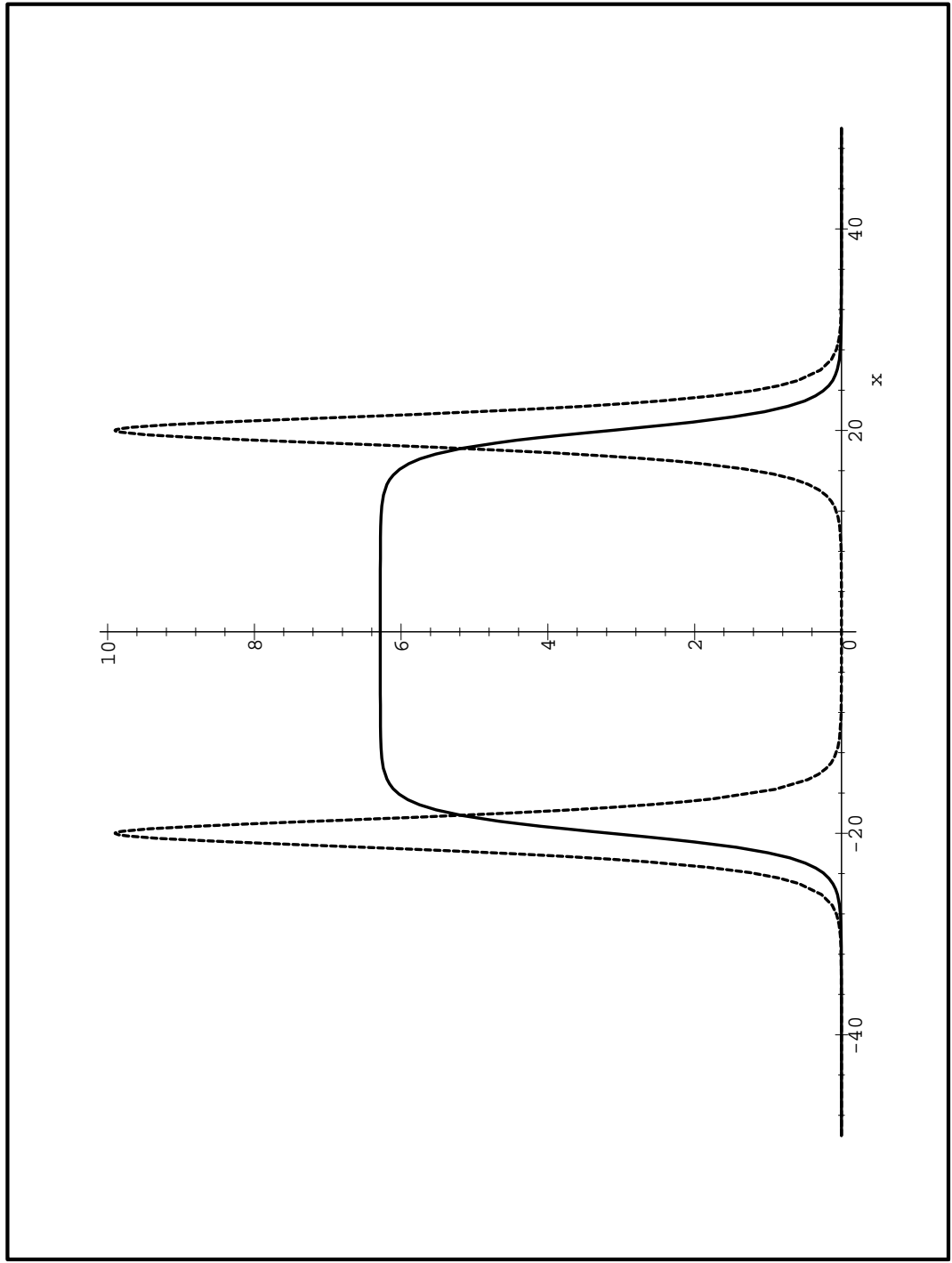}
\caption{Graph of soliton-anti-soliton scattering solution (solid line)
 and corresponding
dilaton(dashed line) for fixed t (not to scale).}
\label{sol_antisol}
\end{figure}

It is useful to write \req{2-sol1} in a somewhat different form.
We proceed by writing $\ell^2:=e^\sigma$ and factoring out an 
$\epsilon_1$ from the numerator and $\epsilon_1\epsilon_2 e^\sigma$ from 
the denominator in the argument of the inverse tangent in \req{2-sol1}.  
Then after multiplying the numerator and denominator by $\exp[\half(\sigma 
-\rho_1-\rho_2)]$ we obtain
\beq
u=4\tan^{-1}{\epsilon_2\over\ell}\left[{e^{\half(\rho_1-\rho_2)}+
\epsilon_1\epsilon_2 e^{-\half(\rho_1-\rho_2)}\over e^{\half(\rho_1+\rho_2
+\sigma)}+\epsilon_1\epsilon_2 e^{-\half(\rho_1+\rho_2+\sigma)}}\right].
\eeq
 Now choose 
new parameters $\mu,v$ in terms of the $v_1,v_2$ by solving the equations 
\bea
\gamma_1-\gamma_2&=&2\gamma v\sin\mu,\\
\gamma_1+\gamma_2&=&2\gamma\cos\mu,\\
\gamma_1 v_1-\gamma_2 v_2&=&-2\gamma\sin\mu,\\
\gamma_1 v_1+\gamma_2 v_2&=&2\gamma v\cos\mu.
\eea
In terms of the new parameters, $l=\tan\mu$ and we obtain:
\beq
u=4\tan^{-1}{(F/G)},\label{2sol}
\eeq
where for the $N=2$ soliton
\bea
F&:=&\cot{(\mu)}\sinh{[m\sin{(\mu)}\gamma(t+vx)]},
\label{F}\\
G&:=&\sinh{[m\cos{(\mu)}\gamma(x-vt)]},
\label{G}
\eea
whereas for the soliton-anti-soliton
\bea
F&:=&\cot{(\mu)}\cosh{[m\sin{(\mu)}\gamma(t+vx)]},
\label{Fc}\\
G&:=&\cosh{[m\cos{(\mu)}\gamma(x-vt)]},
\label{Gc}
\eea
Note that we have absorbed terms of $\half\sigma$ in the exponents into the 
parameters $w_1,w_2$, without loss of generality. In Figs. \ref{2soliton} and 
\ref{sol_antisol}, the
N=2 soliton and soliton-anti-soliton solution are graphed for fixed $t$.

For a more complete description of the N-soliton solutions,
 see the review articles in \cite{early} 
or \cite{ceg}. 

\section{Black Hole Geometries from Multi-Solitons}

We now display the explicit black hole geometries associated with the 
1- and 2- soliton solutions of the sine-Gordon equation.
The  1-soliton solution of the `euclidean' sine-Gordon equation  
can be written
\beq
u(t,x)=4\tan^{-1}\exp\left\{\pm m\gamma(x-vt-\delta_0)\right\},\label{1sol}
\eeq
with $\gamma:=(1+v^2)^{-\half}$,
and $\delta_0=w/m\gamma$ is an 
integration constant.  The constant $v$ is a `spectral parameter'.  
The solution with the $+$ sign in the exponent is the 1-soliton
solution; the opposite sign is the anti-soliton solution.
\footnote{It seems that the $\pm$ sign determining the solitonic/
anti-solitonic nature of the solution has migrated from a factor 
multiplying the exponential function in \req{1sol_1} into the exponential 
itself in \req{1sol}.  In fact the solutions differ by $2\pi$, and so are 
equivalent.} 
  Upon `Wick
rotation' to the Lorentzian signature, (and in this case $v\to iv$), one
sees that the soliton(anti-soliton) propogates through space with constant
velocity $v$ $(-v)$.  Hence we may think of the soliton as being located at
$x=vt$ at time $t$.

We shall now demonstrate that the 1-soliton solution \req{1sol} of the
sine-Gordon equation determines a metric in a coordinate patch on $M_2$
in which there is a Killing vector field which is timelike in the
region outside the event horizon, but which becomes null 
at an interior point of the patch.  In
other words, it determines a black hole metric.  Indeed, when \req{1sol} is
used in the Lorentzian metric \req{eq: adm}, the latter simplifies to:
\beq
ds^2_{1-sol}=- \hbox{sech}^2{\rho} dt^2+\tanh^2{\rho} dx^2,\label{1sGbh}
\eeq
where
\beq
\rho:=m\gamma(x-vt),
\eeq
and we have chosen for simplicity $\delta_0=0$.  

According to the analysis in Section 3, we may transform to black 
hole coordinates $(T,r)$ if we have a solution $\tau$ to the dilaton
equations for metric given by \req{1sGbh}. Such a dilaton can easily
be found by recalling \cite{soliton1} that the dilaton equations imply
that the field $\tau$ also satisfies the linearized sine-Gordon 
equation \req{lsg}.
 It is
straightforward to show 
that the linearized sine-Gordon equation is automatically
satisfied by a field of the form:
\beq
\tau = a \dot{u} + bu',
\label{dilansatz}
\eeq
where
$a$ and $b$ are arbitrary constants. We therefore take \req{dilansatz}
as our {\it anasatz} and then see whether there are values for $a$ and 
$b$ for which the remaining dilaton equations are satisfied. In the 
one soliton solution \req{1sol}
\beq
\dot{u}=\mp 4 m\gamma v \hbox{sech}\rho = -v u',
\eeq
and  \req{dilansatz}
satisfies all the dilaton equations for any $a,b$. In the above, the minus
and plus signs refer to the soliton and anti-soliton respectively.
We therefore choose $b=0$, so that:
\beq
\tau= 4m|av| \gamma \hbox{sech}(\rho),
\label{dil_1sol}
\eeq
where we assume that the sign of $a$ has been chosen to make $\tau$
positive.
The black hole coordinates $(r,T)$ can therefore be defined
by
\bea
r&=&\tau/m={4|av|\gamma }\hbox{sech}{\rho},\\
dT&=&(4|av|m^2\gamma^2)^{-1}\left[dt-v{\tanh^2{\rho}\over m\gamma(\hbox{sech}^2{\rho}-
v^2\tanh^2{\rho})}d\rho\right],
\label{1soltrans}
\eea
In these coordinates, the metric is of the form
\beq
ds^2_{bh}=-\left(m^2 r^2-16m^2 a^2\gamma^4 v^4\right) dT^2+\left(m^2 r^2-
16m^2 a^2\gamma^4 v^4\right)^{-1}dr^2.
\label{1bh}
\eeq
This is the metric of a Jackiw-Teitelboim black hole with mass parameter
\beq
M_{1sol} = 4am\gamma^2 v^2,
\eeq and event horizon at $\tau=
\tau_H=4m|a|\gamma^2 v^2$.
It is important to note that the mass $M$ is non-negative for all
values of $v$ and $a$.
The choice of the normalization constant $a$  is discussed
 in the following  
section on the Hamiltonian analysis.  

\begin{figure}[hbt]
\vspace*{-3cm}
\begin{center}
\leavevmode
\epsfxsize=16 cm
\epsfbox{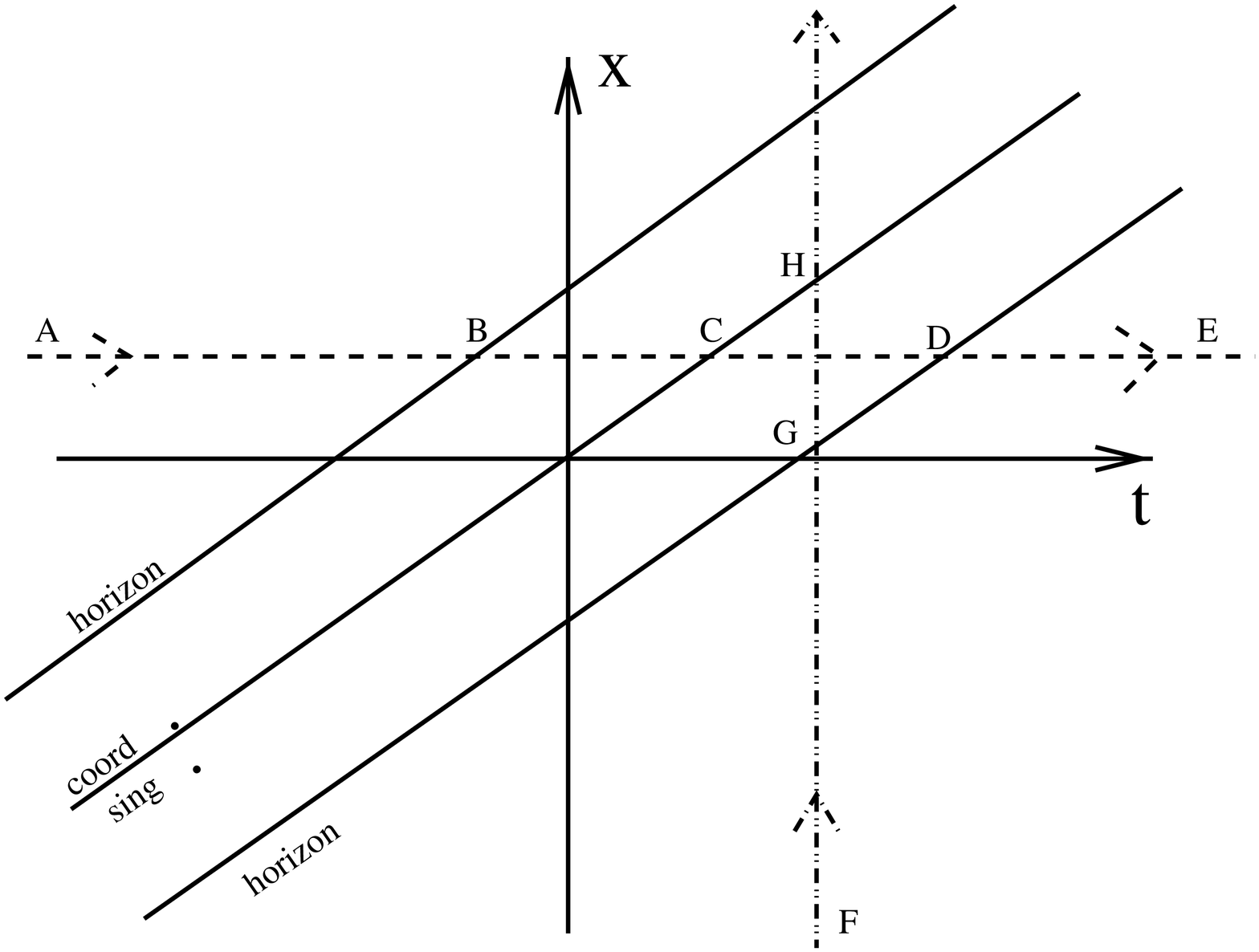}
\end{center}
\caption{Horizons and coordinate singularity in 1-soliton coordinates.}
\label{1soliton}
\end{figure}

\begin{figure}[hbt]
\vspace*{-3cm}
\begin{center}
\leavevmode
\epsfxsize=16 cm
\epsfbox{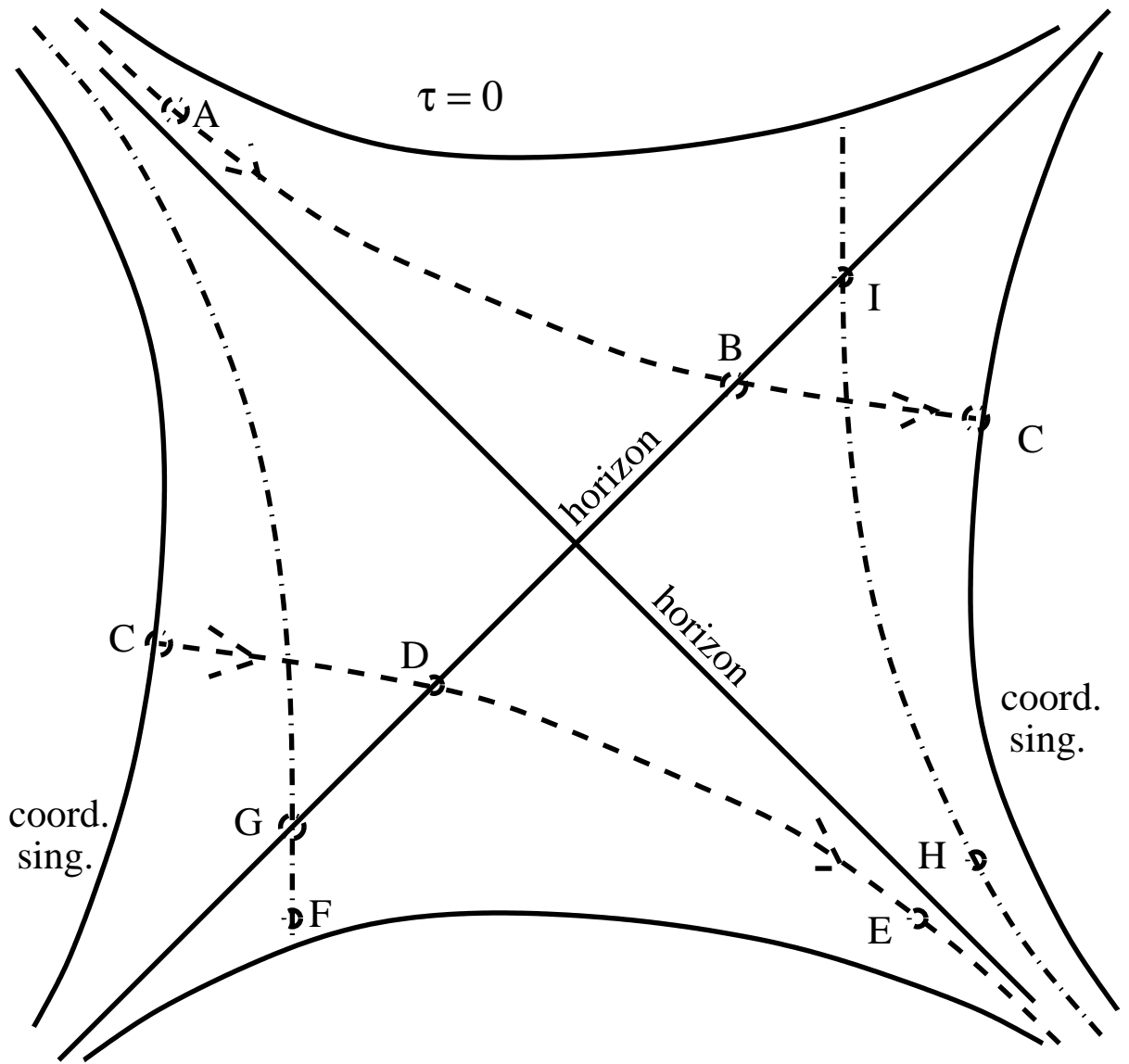}
\end{center}
\caption{Surfaces of constant x (A..B..C..E) and t (F..G..H..I)
in Kruskal coordinates.}
\label{kruskal}
\end{figure}

As noted previously, the sine-Gordon metric \req{1sGbh} is Kruskal-like in that there is
no coordinate singularity at the horizon.  The metric is regular on a 
patch extending from $\rho=-\infty$, where $\tau=0$, to the location of
the soliton, where $\rho=0$, 
where $\tau=\tau_C=4ma\gamma v$. Thus the location of the 
sine-Gordon soliton is the surface
along which the sine-Gordon coordinates break down. Since the ratio:
\beq
{\tau_H\over\tau_C}= \gamma v= {v\over\sqrt{1+v^2}},
\eeq
the soliton is always located outside the horizon. The sine-Gordon
coordinates are therefore regular at the horizon. Moroever, by taking the
limit $v\to\infty$, we can place the soliton arbitrarily close to the 
horizon.  We note here that the metric corresponding to the 1-soliton in 
the limit as $v\to\infty$ is
\beq
ds^2=-\hbox{sech}^2(mt)dt^2+\tanh^2(mt)dx^2. \label{infinitev}
\eeq 
This metric has constant curvature $-2m^2$.  The corresponding 
mass parameter $M^\infty=4|a|m$ and now the location of the soliton, at 
$t=0$, coincides with $\tau^\infty=0$, where the dilaton $\tau^\infty$ 
is 
\beq
\tau^\infty={1\over m}\cosh(mx)\,tanh(mt).\label{tauinf}
\eeq
Hence the entire spacetime, excluding $\tau^\infty=0$, but including 
the asymptotic region $r=\tau^\infty/m\to\infty$, is covered by the 
sine-Gordon coordinate patch.  
Fig. \ref{1soliton} 
illustrates the locations of the event horizons and coordinate
singularites in 1-soliton sine-Gordon coordinates. Fig.\ref{kruskal}
shows how a generic surface of constant soliton coordinates $x$ 
(A..B..C..D..E)  and $t$ (F..G..H..I) are 
 embedded in the Kruskal diagram for  the corresponding
black hole. Note that both $\tau=0$ and $\tau=\tau_C$ are
clearly coordinate singularites in the soliton coordinates, since they
are reached only asymptotically by lines of constant $x$ and 
$t$, respectively.

We now discuss the 2-soliton coordinates.
The metric, in this case given by \req{2sol}, is
\beq
ds^2_{2-sol}=-2{FG\over F^2+G^2}dt^2+{G^2-F^2\over F^2+G^2}dx^2,
\eeq
where the quantities $F$ and $G$ are given by either \req{F} and \req{G} 
or \req{Fc} and \req{Gc} above.

\begin{figure}[hbt]
\vspace*{-3cm}
\leavevmode
\epsfxsize=16 cm
\epsfbox{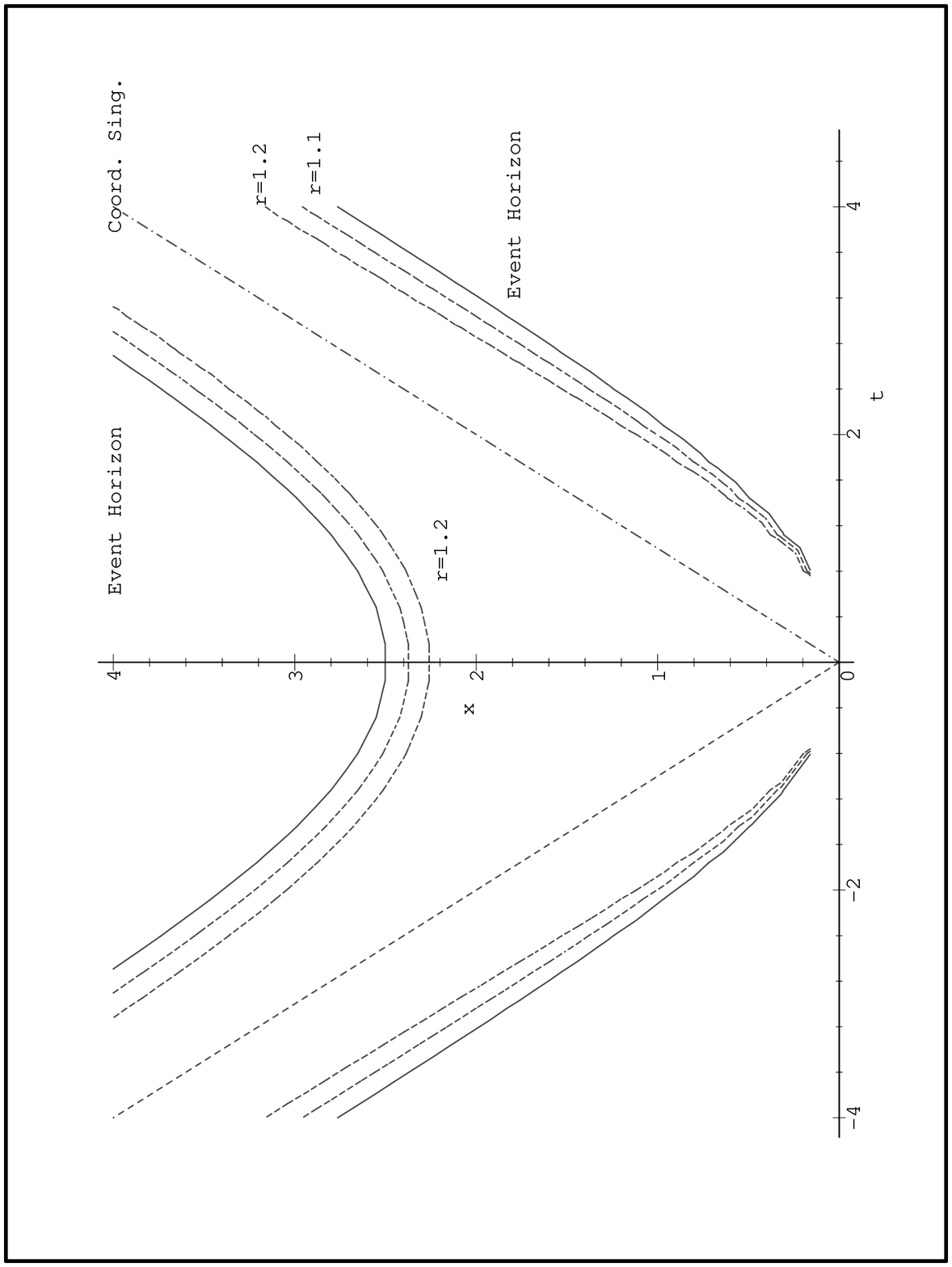}
\caption{Horizons and coordinate singularity in 2-soliton coordinates.}
\label{2sol_hor}
\end{figure}

\begin{figure}[hbt]
\vspace*{-3cm}
\leavevmode
\epsfxsize=16 cm
\epsfbox{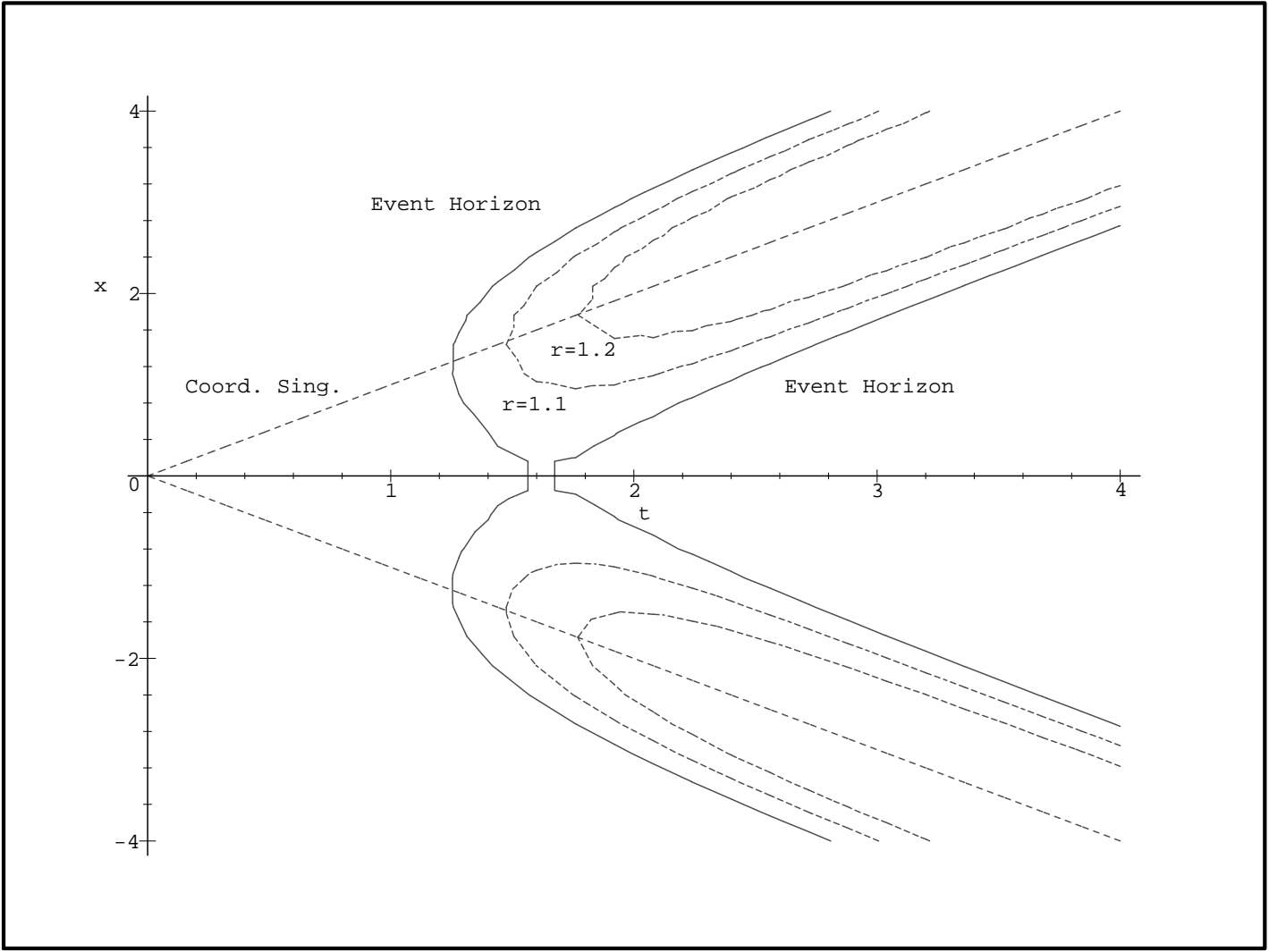}
\caption{Horizons and coordinate singularities in soliton-anti-soliton
coordinates.}
\label{sol_antisol_hor}
\end{figure}

Using MAPLE, we computed the dilaton for the 2-soliton metric above
by invoking that {\it ansatz} \req{dilansatz}. It turns out that 
this  {\it ansatz}
satisfies all three dilaton equations providing that 
$b = 2 v a/(1-v^2)$. The resulting dilaton, for the case where $F,G$ are 
given by \req{F},\req{G}, is:
\beq
\tau={4am\cot\mu\over\gamma(v^2-1)}{\left[v\cos\mu\cosh{\rho_-}\sinh{\rho_+}
-\sin\mu\cosh{\rho_+}\sinh{\rho_-}\right]\over\left[\sinh^2{\rho_-}+
\cot^2{\mu}\sinh^2{\rho_+}\right]},\label{2tau}
\eeq
where $\rho_+:=m\gamma(t+vx)\sin\mu,\rho_-:=m\gamma(x-vt)\cos\mu$.
The corresponding conserved mass parameter is:
\beq
M_{2sol}=\left[{2am\left(v^2\cos^2\mu-\sin^2\mu\right)\over v^2-1}\right]^2.\label{M2}
\eeq
It is interesting that this is again non-negative for all values
of the soliton parameters.
See Fig.\ref{2sol_hor} for the structure of the horizons, coordinate singularities and 
some constant $\tau$ curves for the geometry in these coordinates.  For the 
soliton-anti-soliton scattering solution, i.e. the
case where $F,G$ are given by \req{Fc},\req{Gc}, the expression for the 
dilaton is given by \req{2tau} but with $\sinh$ and $\cosh$ interchanged;  
while the expression for the conserved mass parameter is identical to \req{M2} 
above.  Fig.\ref{sol_antisol_hor} displays some of the geometrical features.

\section{Hamiltonian Analysis}
\medskip
We now review the Hamiltonian analysis for Jackiw-Teitelboim 
gravity, using the notation of \cite{obs}.
Spacetime is split into
a product of space and time:  $M_2 \simeq
\Sigma\times
R$ and the metric $h_{\mu\nu}$ is given an ADM-like
parameterization:\cite{torre}
\beq
ds^2=e^\alpha\left[-\sigma^2dt^2+\left(dx+Vdt\right)
^2\right]
.\label{eq: adm}
\eeq
where $\alpha$, $V$
and
$\sigma$ are functions on spacetime $M_2$.
In
the following, we denote by the overdot and prime, respectively,
derivatives with respect to the time coordinate $t$ and spatial
coordinate $x$.
\par
The canonical momenta conjugate to the fields
$\{\alpha,\tau\}$ are:
\bea
\Pi_\alpha&=&{1\over2G\sigma}\left(V\tau'-\dot\tau\right),
\label{eq: pi_alpha}
\\
\Pi_\tau&=&{1\over2G\sigma}\left(-\dot\alpha+V\alpha'+2V'
\right),
\label{eq: pi_tau}
\eea
The vanishing
of the momenta canonically conjugate to $V$ and $\sigma$
yield the primary constraints for the system. Following the
standard
Dirac prescription\cite{dirac}, we obtain the canonical
Hamiltonian (up to spatial divergences):
\beq
H_0=\int dx\left(V{\cal F}+{1\over2G}\sigma{\cal
G}\right).\label{eq: ham}
\eeq
where we have defined:
\beq {\cal F}:=\alpha' \Pi_\alpha+\tau' \Pi_\tau-2\Pi_\alpha',
\label{eq: gauss}
 \eeq
\beq
{\cal
G}:=2\tau''-\alpha'\tau'-\left(2G\right)^2
\Pi_\alpha
\Pi_\tau-2m^2{e^\alpha}\tau.\label{eq: hamcon}
\eeq
Clearly ${1\over2G}\sigma$ and $V$ play the role
of Lagrange multipliers that enforce the secondary constraints
${\cal F}\approx 0$ and ${\cal G}\approx 0$.
\par
The energy can be constructed by
noting that the
following linear combination of the constraints is a total spatial
derivative:
\bea
\tilde{\cal G}&:=&{l\over 2} e^{-\alpha}\left((2G)^2\Pi_\alpha{\cal
F}+
\tau'{\cal G}\right)\none
&=&(q[\alpha,\tau,\Pi_\alpha,\Pi_\tau])'\\
&\approx& 0  ,
\label{eq: def'n q'}
\eea
where we have defined the variable $q$ as
\beq
q:={1\over 2m}\left[e^{-\alpha}\left((2G\Pi_\alpha)^2-
(\tau')^2\right)+ m^2\tau^2\right].\label{eq:qdef}
\eeq
The expression on the right-hand side above is nominally an
implicit function
of the spatial
coordinate, but is constant on the constraint surface. Moreover,
it is straightforward to show
that $q$ commutes with both
constraints ${\cal F, G}$. Thus, the constant mode of $q$ is a
physical
observable in the Dirac sense.
\par
In terms of the canonical momenta the magnitude of the Killing
vector can be written as:
\beq \mid
k\mid^2={e^{-\alpha}\over m^2}\left[(2G\Pi_\alpha)^2-(\tau')^2\right] \,\,
{}.
\label{eq:kill/mom} 
\eeq
Thus the observable $q$ is:
\bea
q&=&{m\over 2}\left(\mid k\mid^2+ \tau^2\right)\none
  &=& {M m\over 2} \,\, .
\label{eq:cov q}
\eea
\par
The momentum conjugate to $q$, is found by inspection to
be\cite{domingo1}:
\beq
p:=-\int_\Sigma dx {2\Pi_\alpha e^\alpha\over
(2G\Pi_\alpha)^2-(\tau')^2} \,\, .
\label{eq:pdef}
\eeq
The value of $p$ depends on the global properties of the
spacetime slicing. This is consistent with the generalized Birkhoff
theorem\cite{domingo1} which states that there is only one
independent diffeomorphism invariant parameter
characterizing the space of solutions.
\par
It is instructive to write the observable $p$ in covariant form:
\bea
p&=&\int_\Sigma dx e^{\alpha/2}n^\mu{\nabla_\mu\tau\over\mid
k\mid^2}\\
  &=&-2\int_\Sigma dx^\mu {k_\mu\over\mid
k\mid^2}\\
&=& -2m\int_\sigma dT .\label{eq: cov p}
\eea
Note that $dxe^{\alpha/2}$ is the measure induced on $\Sigma$ by
$h_{\mu\nu}$.
In the expression for $p$ the vector field $n^\mu$ is the unit
(timelike)
normal to $\Sigma$. The final expression is obtained by using
the result \req{dT}, and proves explicitly that the momentum
conjugate to $M$ is equal to the ``Schwarzschild time separation'' of
the slice\cite{kuchar,obs}.
\par
The
canonical Hamiltonian in terms of $\tilde{\cal G}$ is:
\beq
H_0 =\int dx \left(\tilde{v} {\cal F}
   -\tilde{\sigma}{q'\over G}\right) + H_+-H_- \,\,
{}.
\label{eq: ham2}
\eeq
where $\tilde{v}= V- 2G\sigma \Pi_\alpha/\tau'$ and
\beq
\tilde{\sigma}={m\sigma e^\alpha \over  \tau'}.
\eeq
Note that from \req{eq: adm} it follows that 
\beq
\sigma e^\alpha = \sqrt{|g|} = |\sin(u/2)\cos(u/2)|
\eeq
where the last expression is only valid in soliton coordinates.
 In \req{eq: ham2} $H_+$ and 
$H_-$ are surface terms needed to make the variational principle well defined.
These surface terms depend on the boundary conditions, and will be determined
below.
\par
We now impose boundary conditions on our spatial slice consistent with
soliton coordinates \req{sGmetric}. In particular, we assume that the spatial
coordinate $x$ runs form $X_-=-\infty$ to $X_+=+\infty$. At the inner boundary
$X_-$ the metric and dilaton should take on values corresponding to 
the asymptotic ($x\to-\infty$) region of a constant $t$ surface in
soliton coordinates. As illustrated for the one-soliton case in
Fig.\ref{kruskal}, such surfaces approach $\tau=0$ asymptotically
along the horizon.\footnote{Slicings of this general form were
  considered
for spherically symmetric gravity in \cite{bose}.}
Thus, we require $V_-\to0$, $\sigma_-\to 0$, $\Pi_\tau|_-\to 0$,
 $e^\alpha|_-\to 1$,
 $\tau_-\to0$ and $\tau'_-\to 0$. However, in order for the Hamiltonian to be
well defined, $\tilde{\sigma}$ must be finite at the boundary, so we restrict
$\tilde{\sigma}_- =constant$. This condition has two important consequences. 
First
it allows the boundary terms to be integrated in a straightforward fashion,
as shown below. Secondly, once the soliton metric is specified, it fixes the
scale of the dilaton. That is,
given any soliton solution, there exists a corresponding black hole
with uniquely determined mass.\footnote{Another way to state this is that each soliton 
solution $u$ provides a unique slicing of the interior of 
black hole spacetime of fixed
mass.}
As we saw in Section 5, without this condition the linearity of the dilaton equations
of motion allow an arbitrary multiplicative scale factor in the solution for
the dilaton, and the resulting black hole mass observable is 
proportional to the square of 
the scale factor. However, in order to be able to impose this boundary condition on
$\tilde{\sigma}$ it is necessary that $|\sin(u/2)\cos(u/2)|/\tau'$ remain 
finite as $x\to-\infty$ for every soliton solution $u$ and dilaton field
$\tau$. We have been able to verify this explicitly in the 1- and 2- soliton
sector, but not in the general case. Specifically, in the one soliton case,
for the solution given by \req{1sol} and \req{dil_1sol}
we find that $|\sin(u/2)\cos(u/2)|/\tau'= 1/(4m\gamma^2)$
 for all $x$, so we choose
$a=1/(4m\gamma^2)$ and the corresponding black hole mass is $M =  v^2$ with
corresponding ADM energy $E=mv^2/(2G)$. It is interesting that the ADM
energy of the black hole is equal to the (non-relativistic)
kinetic energy of the soliton. 
\par
In the
two soliton case, one finds that 
\beq
|\sin(u/2)\cos(u/2)|/\tau'\to  
\left({v^2-1\over 2 a}\right)
\left[{|v\sin\mu\cos\mu|\over v}(1+v^2)\right]^{-1},
\eeq
as $x\to-\infty$, so we choose 
\beq
a=\left({v^2-1\over 2 }\right)
\left[{|v\sin\mu\cos\mu|\over v}(1+v^2)\right]^{-1},
\eeq 
to get a mass:
\beq
M=v^2\left[\cos^2\mu v^2-\sin^2\mu\over
v^2-|v\sin\mu\cos\mu|(1+v^2)\right].
\eeq
\par
The choice of boundary conditions at the outer boundary is somewhat more 
delicate. In order to consider black hole dynamics and thermodynamics we 
would like our spatial slice to include the asymptotic region of the black hole. Soliton coordinates, as discussed above, cannot be extended into the
asymptotic region since there is a coordinate singularity
 at $u= (2n+1)\pi/2$, which  corresponds to the location of a soliton. We
avoid this problem by assuming 
 that at $X_+\to \infty$, our spatial slice approaches asymptotically
a static Schwarzschild slice, with no coordinate singularity  between
$X_-$ and $X_+$. This requires a change of coordinates between the horizon
and the soliton location, since soliton coordinates are good in the neighbourhood of the horizon, while Schwarzschild coordinates are good in the 
neighbourhood of the soliton. As discussed in
\cite{obs}, the only  boundary conditions that we require at $X_+$ are 
$\tilde{\sigma}_+\to 1$, $\tilde{v}_+\to 0$. \footnote{For a detailed Hamlitonian analysis with exterior boundary conditions corresponding to a black hole
in a static box, see \cite{kunst}.}

\par
Given the above boundary conditions it is possible
to evaluate the surface terms for any solitonic solution of the 
sine-Gordon equations. Using the
identity:
\beq
\alpha'\Pi_\alpha-2\Pi_\alpha'=- {e^\alpha\over\Pi_\alpha}
(e^{-\alpha}\Pi_\alpha^2)',
\eeq
we first write the canonical Hamiltonian in the following form:
\beq
H_0 =\int dx \left(-{\tilde{v}e^\alpha\over \Pi_\alpha}
 (e^{-\alpha}\Pi_\alpha^2)' +\tilde{v}\Pi_\tau\tau'
   -\tilde{\sigma}{q'\over G}\right) + H_+-H_- .
\eeq
The variation of $H_0$ contains the following  boundary terms:
\beq
\delta \left.H_0\right.|_{\hbox{boundary}}
 = \int dx \left(-{\tilde{v}e^\alpha\over \Pi_\alpha}
\delta (e^{-\alpha}\Pi_\alpha^2) +\tilde{v}\Pi_\tau\delta\tau
   -\tilde{\sigma}{\delta q\over G}\right)' +\delta H_+-\delta H_- .
 \eeq
Given the above boundary conditions, potentially non-zero contributions are:
\beq
\delta H_0|_{\hbox{boundary}}
 = -\tilde{\sigma}\left.{\delta q\over G}\right|_+ -
\left.\left({\tilde{v}e^\alpha\over \Pi_\alpha}
\delta (e^{-\alpha}\Pi_\alpha^2)
   -\tilde{\sigma}{\delta q\over G}\right)\right|_- +\delta H_+-\delta H_- .
 \eeq
Using the expression for $q$ with for $\tau=\tau'=0$, and the fact that
when $V=0$, ${\tilde{v}e^\alpha\over \Pi_\alpha}= -\tilde{\sigma}$, we
find that the there is no surface contribution at $X_-$, whereas the 
surface contribution in the asymptotic region can easily be integrated to
give:
\beq
H_+ = q/G = {M m\over 2G} .
\eeq

\section{Conclusions}
We have discussed in some detail how Euclidean sine-Gordon solitons
can be used to coordinatize black holes in Jackiw-Teitelboim gravity.
The solitons appear as coordinate singularities that constitute the boundaries
of the patches that can be faithfully coordinatized by the sine-Gordon 
coordinates. The horizons generically are regular in these coordinates.
In the one-soliton case the soliton was a surface of constant dilaton
field that lay just outside the horizon. There are still many unanswered questions about how our specific results for the 1- and 2- soliton sectors
generalize to the N-soliton case.

Of course the most important question concerns whether or not there
is any physics in this. It is tantalizing to speculate on what would
happen if we were able to treat the solitons as physical particles
propagating through the black hole spacetime, and providing a
physical boundary whose deformations are in some way related to the
diffeomorphisms of the horizon itself. Since in Carlip's program
\cite{carlip} the diffeomorphisms of the horizon may be related to
the black hole entropy, we might be able to quantize the horizon
diffeos by quantizing the solitons and account for the black hole
entropy by counting soliton states.

The following is evidence that such a proceedure may be worth 
pursuing.  
A black hole state has total energy $E_0$.  Suppose it is described 
by an N-soliton solution of the sine-Gordon equation.   
(Ignore breathers and other exotica for now.)  
Ignoring breathers, etc., Takhtadjan and Faddeev \cite{tf} compute 
the total energy, valid both classically and quantum mechanically,  
\beq
E=\sum_i^N\left(64m^2/\beta^{2}+p^2_i\right)^{1/2},
\eeq
where $p_i$ is canonical momentum of $i^{th}$ 
lump and $\beta$ is the sine-Gordon coupling constant.  
Absorb factors of $64$ into 
$\beta^{-2}$, so the rest energy of the state is $E_0=mN/\beta$.
Now combinatorics come in.  The degeneracy of the state 
arises from the indistinguishability of the  
lumps in the N-soliton state.   In other words, the degeneracy is the 
number of different ways to write N as the sum of non-negative integers.  
This is 
the number-theoretic partition function of Hardy-Ramanujan \cite{hr}
\beq
n(N)\sim e^{\pi\sqrt{2N/3}},
\eeq
for large $N$.  Hence the entropy behaves as 
\beq
S\sim\log{n(N)}\sim \pi\sqrt{N}\sim\pi\sqrt{E_0}.
\eeq
This is just the Bekenstein-Hawking entropy (up to factors of order 1) 
for a Jackiw-Teitelboim black hole of 
with total energy $E_0$.

This is quite sketchy, as well as speculative.  In order to make the
argument more rigorous, at least the following must be addressed:\\
(1.)  Can one ignore breathers and other non-solitonic solutions of
the sine-Gordon equation?   \\
(2.)   It is not obvious that the black hole energy is given by the rest 
energy of the N-soliton solution.  It is true however that the energy of the 
black hole corresponding to a 1-soliton in the limit as the soliton
parameter $v\to\infty$ is (up to numerical factors of order unity) the 
same as that of the 1-soliton itself.

Work is in progress to address these issues.

\bigskip
\par\noindent
{\large\bf Acknowledgements}
\par
The authors would like to thank Y. Billig and M. Ryan for useful conversations.  This work
was supported in part by the Natural Sciences and Engineering
Research
Council of Canada. G.K is grateful to M. Ryan and A.A. Minzoni for
helpful conversations.
 \par
\newpage

\end{document}